\documentclass[fleqn,twoside,twocolumn,nofootinbib]{revtex4} 
\usepackage[sec]{ujp} 
\begin{document}
\title[ON THE POSSIBILITY OF \textquotedblleft DRY\textquotedblright\
FRICTION]
{ON THE POSSIBILITY OF \textquotedblleft DRY\textquotedblright\
FRICTION\\ IN SUPERFLUID He\boldmath$^{4}$}%
\author{M.D. Tomchenko}
\affiliation{\bitp}
\address{\bitpaddr}
\email{mtomchenko@bitp.kiev.ua}
\udk{538.941} \pacs{67.10.Hk; 67.30.hp;\\[-3pt] 68.08.-p} \razd{\secvi}

\setcounter{page}{144}%
\maketitle


\begin{abstract}
A possible microscopic explanation for the exhaustion of $\rho_{s}$
in helium-II at the wall at $T>T_{c}\simeq0.5\div1$\,K has been
proposed, and a possibility for the \textquotedblleft
dry\textquotedblright\ friction to exist in He-II at $T\leq T_{c}$
has been predicted. Both effects are related to the fact that the
energy of 2D-rotons is lower by 2~K than that of 3D-rotons, so that
the wall is a potential well for the latter.
\end{abstract}

\vskip1cm \noindent At wetting, helium atoms stick to a wall.
However, taking into account that $\mathrm{rot~}\mathbf{v}_{s}=0$,
the velocity $\mathbf{v}_{s}$ cannot grow permanently, when moving
away from the wall. From this reason, V.L.~Ginzburg drew
conclusion \cite{ginz1} that $\mathbf{v}_{s}$ has a jump near the
wall
and, therefore, there must be \textquotedblleft dry\textquotedblright%
\ friction in He~II. However, such friction was not found experimentally
\cite{gam}, whence it follows \cite{ginz2} that $\rho_{s}=0$ at the wall.
Later, this hypothesis was confirmed experimentally \cite{pat}. To our
knowledge, the microscopic reason of such an \textquotedblleft
exhaustion\textquotedblright\ of $\rho_{s}$ has not been elucidated yet.
In addition, the helium temperature was not specified in work \cite{gam}.
However, if ultralow temperatures are not required, experiments with helium-II
are usually carried out at $T>1.2~$K, which is associated with a cooling
procedure. Below in this work, simple microscopic reasonings are proposed,
which can explain the exhaustion of $\rho_{s}$ at the wall and allow us to
predict \textquotedblleft dry\textquotedblright\ friction at $T\lesssim
1~\mathrm{K}$.

By definition, $\rho_{s}=\rho-\rho_{n}$. Therefore, the exhaustion
of $\rho_{s}$ at the wall can originate from the behavior of either
$\rho$ or $\rho_{n}$. It is known that the $\rho$ corresponds to the
atomic density, and $\rho_{n}$ does to the quasiparticle one. The
properties of $\rho$ can ensure the equality $\rho_{s}=0$ in two
cases. First, it can be, if $\rho=0$. However, the exact zero is
impossible, because the wall is not an infinitely high energy
barrier, so that the wave function of He~II, together with $\rho$,
must be different from zero. Second, it can take place as a result
of the exact equality $\rho=\rho_{n}$. However, nothing forces atoms
to arrange in such a manner that the equality $\rho=\rho_{n}$ be
satisfied just at the wall. The following variant is also possible:
the $\rho$-value is very low near the wall, so that $\rho_{s}$ is
also small; in this case, dry friction does exist, but it is too low
to be detected. However, in this case, $\rho_{s}$ must be close to
zero only in a close vicinity to the wall, at distances not farther
than the average interatomic one (because nothing prevents atoms
from approaching so closely). However, in accordance with the
experiment \cite{pat,3sound1}, $\rho_{s}$ is close to zero at larger
distances from the wall (approximately 2 atomic layers). Whence it
follows that it is more likely the properties of $\rho_{n}$ rather
than $\rho$ that are responsible for the equality $\rho_{s}=0$ at
the wall. In other words, $\rho_{s}\rightarrow0$ near the wall as a
result of the quasiparticle behavior rather than the atomic one:
owing to a certain reason, the highest possible concentration of
quasiparticles is attained at the wall, and the condition for
$\lambda $-transition, $\rho_{s}=0$, is realized. Let us examine
this variant. The fact that the thickness of a helium layer, for
which $\rho_{s}\approx0$ (about 2 atomic layers), approximately
coincides with the effective radius of a roton \cite{rr} (about
1.5~atomic layers) testifies in favor of this hypothesis.

The following simple mechanism is possible. From the results of
microscopic calculations \cite{krot1,krot2} and the experiment
\cite{2dr}, it follows that the energy $\Delta_{\mathrm{2D}}$ of a
surface (2D) roton is lower by approximately 2$~\mathrm{K}$ than the
energy of a bulk (3D) roton. From the dispersion curves for 2D- and
3D-rotons (see Figure), it is evident that two processes may run
near the wall: (\textit{a}) a direct one, i.e. a 3D-roton creates a
2D-roton and a 3D-phonon, and (\textit{b}) an inverse one: a
2D-roton and a 3D-phonon merge to create a 3D-roton. One may choose
an arbitrary point in the 3D-curve in a vicinity of the roton
minimum and draw two straight lines from it downwards to the right
and to the left at a definite angle with respect to the vertical,
which is equal to the slope angle of the phonon branch. The
intersection points of those straight lines with the 2D-roton curve
determine possible states of a 2D-roton, whereas the vector
connecting those points determines a required 3D-phonon. In this
case, the conservation laws for energy and momentum can be
satisfied. The only restriction is that the $z$-component of the
3D-roton momentum has to be small. The binary process 2D-roton
$\rightarrow$ 3D-roton and the inverse one, as well as the ternary
process 2D-roton $\rightarrow$ 3D-roton + 3D-phonon and the inverse
one, are forbidden by the conservation laws. There may also be
quaternary and higher-order processes, but their probabilities are
low.

Therefore, two processes, (\textit{a}) and (\textit{b}), dominate
among the roton-assisted ones running near the wall. Every
3D-roton that approaches the wall can decay into a 2D-roton and a
3D-phonon with a certain probability. Accordingly, every 2D-roton
can merge, with a certain probability, with a 3D-phonon to create
a 3D-roton. However, those probabilities are evidently different:
the former is governed by the process itself, whereas the latter
is also proportional to the concentration of 3D-phonons with a
momentum that corresponds to the transition. While a 3D-roton can
decay immediately, a 2D-roton has to wait until a required
3D-phonon appears around it. Therefore, the characteristic time of
the latter process has to be larger. As a result, the creation of
a 2D-roton must occur more frequently than its decay. At
$T\gtrsim1~\mathrm{K}$, the number of rotons is large, and those
3D-rotons which come to the wall will decay into 3D-phonons and
2D-rotons, until the highest possible concentration of 2D-rotons
is attained, so that $\rho_{n}$ becomes equal to $\rho$ at the
surface and $\rho_{s}$ vanishes here. It is what is observed in
the experiment. At very low $T\lesssim0.1~$K, the number of rotons
is several orders of magnitude smaller than the number of phonons.
In this case, the equality $T=T_{\lambda}$ at the wall is
impossible. To prove it, let us suppose the contrary, i.e. let the
concentration of 2D-rotons at the wall be maximum, and let
$T=T_{\lambda}$. It is evident that, in this case, process
(\textit{b}) prevails for 2D-rotons, because its probability is
proportional to the considerable concentrations of 2D-rotons and
3D-phonons, whereas the probability of process (\textit{a}) is
proportional to the concentration of 3D-rotons, which is very low
at $T\lesssim0.1~$K. Process (\textit{b}) continues until the
concentration of 2D-rotons reduces to a certain equilibrium value
which is to be calculated. However, even without calculations, it
is clear that the temperature $T$ of the walls is much lower than
$T_{\lambda}$. One may expect that the equilibrium concentration
of 2D-rotons would be of the same order of magnitude as the
concentration of 3D-rotons, i.e. very low. In addition, if one
takes into account that, at low temperatures, phonons with low
energies, for which the dispersion curves for 3D- and 2D-phonons
(the theoretical one) \cite{krot1,krot2} coincide, play a crucial
role, we come to a conclusion that, if the bulk temperature
$T_{\mathrm{3D}}\lesssim0.1~$K, the helium temperature at the wall
has to be close to the former; accordingly, $\rho_{s}\approx\rho$
near the wall.

\begin{figure}
\includegraphics[width=7.5cm]{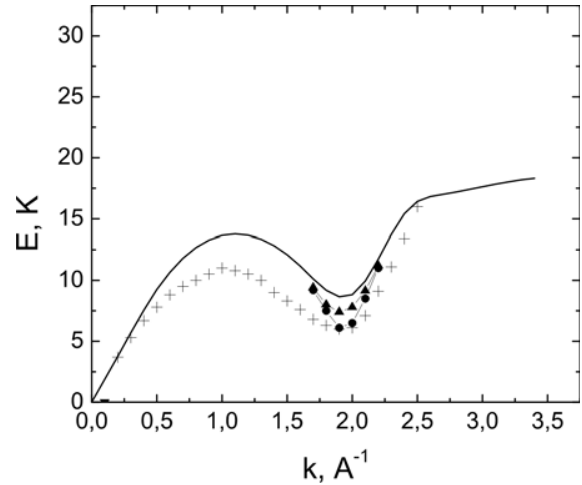}
\vskip-3mm\caption{Experimental dispersion curve for bulk
quasiparticles in He II (solid curve), theoretical dispersion
dependence for surface quasiparticles \cite{krot1,krot2} (crosses),
and experimental dispersion dependences for surface rotons
\cite{2dr} (circles and triangles)  }
\end{figure}

The critical temperature $T_{c}$, at which the exhaustion of
$\rho_{s}$ at the wall disappears, is probably a little lower than
the temperature of transition from the roton domination to the
phonon one and equals to $T_{c}\simeq0.5\div 1~\mathrm{K}$.

Note that our conclusions do not demand any calculations.
It is enough to know that processes (\textit{a}) and
(\textit{b}) run with different rates; therefore, the number of
2D-rotons will either decrease until the wall temperature falls to
the phonon one (of about 1$~\mathrm{K}$) or grow until the
concentration of 2D-rotons at the wall reaches the maximum with
$T=T_{\lambda}$ and $\rho _{s}=0$. The experiment ($\rho_{s}=0$ at
the wall) testifies that the latter scenario is realized. Since the
energy of 2D-rotons is lower by 2$~\mathrm{K}$ than that of
3D-rotons, the wall is a potential well for the latter.

From those reasons, it follows that, in the first helium layers near
the wall, there exists a temperature gradient. From the symmetry
viewpoint, it can emerge as a result of the system isotropy violation
near the wall. This violation is also responsible for the pressure
gradient near the wall \cite{pat}. Note that, as early as in 1941,
P.L.~Kapitsa \cite{kap} observed a temperature jump in a helium
layer $\lesssim0.01~$mm in thickness located near the heater, when
the latter was heated up. This jump was explained in work
\cite{xal-skacokT} as a result of the high thermal conductivity in
helium-II. However, in the calculations of work \cite{xal-skacokT},
the surface excitations of helium were not taken into account.
In accordance with the arguments presented above, those excitations give
rise to a temperature jump in a much thinner helium layer near the
wall, namely, of a few atomic sizes. This jump has a different
nature. It does not require for the heat to be pumped from outside,
and, despite the presence of a temperature gradient, the state is
equilibrium, of course, and the heat flow is absent. Owing to a high
thermal conductivity of helium, the equilibrium is established,
first of all, in the bulk and near-surface helium, whereas the heat
exchange with the wall is much slower. Nevertheless, a weak heat
exchange between surface helium excitations and the wall must
evidently take place. Therefore, there must exist a small jump of
the temperature between the wall and the helium bulk, even in the
absence of a thermal pumping to the wall or helium. It can be verified
in the experiment. In so doing, one has to take into consideration
that such a jump of $T$ must exist both between a heater and
helium, and between a thermometer and helium.

It is of importance that, at $T\leq T_{c}$, $\rho_{s}\neq0$ at the
wall. Therefore, there must exist \textquotedblleft
dry\textquotedblright\ friction \cite{ginz1} which can be tested in
a direct experiment of the type proposed in work \cite{gam} or by
measuring the dependence of the width of the surface roton peak on
the temperature \cite{2dr} (at $T=T_{c}$, the width should jump). In
experiments with the third sound \cite{3sound1}, it was obtained
that the restoration length of $\rho_{s}$ grows with the temperature
at $T\gtrsim1~\mathrm{K}$ and is constant at
$T\lesssim1~\mathrm{K}$. As far as we know, those dependences have
not been explained yet, and the temperature $T\approx1~\mathrm{K}$,
at which the dependence changes its character, may probably be
$T_{c}$.

The properties of the free helium surface are somewhat different
from those of helium near the wall. However, the energy difference
between the surface and bulk rotons is associated, first of all,
with the geometric factor. Therefore, for the free surface, the
equality $\Delta_{\mathrm{2D}}\simeq\Delta _{\mathrm{3D}}-2$ K
should probably be valid as well. In this case, the arguments
presented above are valid, and, at $T_{\mathrm{3D}}\geq T_{c}$,
rotons must condensate on the surface, so that $\rho_{s}=0$ and
$T=T_{\lambda}$ there. In view of a probable high temperature
($\simeq T_{\lambda}$) of the surface helium layer, an issue arises
concerning the temperature of free He atoms above the He~II surface.
For He~II, \textquotedblleft temperature\textquotedblright\ means
\textquotedblleft quasiparticles\textquotedblright. From the
conservation laws, it is evident that, in the case of atoms above
the He~II surface, the probability of the energy exchange with He II
quasiparticles is high for quasiparticles in the bulk and low
(however, nonzero) for near-surface ones. Therefore, the helium
vapor temperature has to be higher than that in the bulk of He II by
a small value, which is to be calculated.

Only the qualitative arguments were presented above. It seems that
the processes in helium-II near the wall and near the free surface
are of interest, being not quite trivial. They remain to be not
clear enough and deserve a more attention from theorists and
experimenters.

\rezume{%
О ВОЗМОЖНОСТИ ``СУХОГО'' ТРЕНИЯ \\В СВЕРХТЕКУЧЕМ He$^4$}{М.Д.
Томченко} {Предложено  возможное микроскопическое объяснение
     истощения $\rho_{s}$ гелия-II
          у  стенки при $T>T_{c} \simeq 0.5 \div 1 \,K$ и предсказана возможность существования в He II ``сухого'' трения при
          $T\leq T_{c}$.
         Оба эффекта связаны с тем, что энергия 2D-ротонов на 2К меньше энергии 3D-ротонов,
     поэтому стенка является потенциальной ямой для последних.}

\end{document}